\documentclass[reprint,showpacs,superscriptaddress,aps,nofootinbib]{revtex4-1}
\usepackage{amssymb}
\usepackage{amsmath,amssymb,graphicx,amsfonts}
\usepackage{graphicx}
\usepackage{dcolumn}
\usepackage{bm}
\usepackage{xcolor,color,colortbl}
\definecolor{Gray}{gray}{0.9}

\def\beq{\begin{equation}}
\def\eeq{\end{equation}}
\def\bea{\begin{eqnarray}}
\def\eea{\end{eqnarray}}

\newcommand{\overbar}[1]{\mkern 1.5mu\overline{\mkern-1.5mu#1\mkern-1.5mu}\mkern 1.5mu}

\usepackage{pstricks}
\usepackage{graphicx}
\usepackage{amsbsy}
\usepackage{bm}
\usepackage{color}
\usepackage{fancyhdr}
\usepackage{epsfig}
\usepackage[colorlinks=true,linkcolor=blue,citecolor=green,urlcolor=blue,filecolor=blue]{hyperref}
\usepackage{slashed}
\usepackage{multirow}
\usepackage{slashed}

\begin{document}

\bigskip

\vspace{2cm}
\title{Exploring GeV-scale Majorana neutrinos in lepton-number-violating $\Lambda_b^0$ baryon decays}
\vskip 6ex

\author{Jhovanny Mej\'{i}a-Guisao}
\email{jmejia@fis.cinvestav.mx}
\affiliation{Departamento de F\'{i}sica, Centro de Investigaci\'{o}n y de Estudios Avanzados del IPN, Apartado Postal 14-740, 07000 Ciudad de M\'{e}xico, M\'{e}xico}
\author{Diego Milan\'{e}s}
\email{diego.milanes@cern.ch}
\affiliation{Departamento de F\'{i}sica, Universidad Nacional de Colombia, C\'{o}digo Postal 11001, Bogot\'{a}, Colombia}
\author{N\'{e}stor Quintero}
\email{nestor.quintero01@usc.edu.co}
\affiliation{Universidad Santiago de Cali, Facultad de Ciencias B\'{a}sicas, Campus Pampalinda, Calle
5 No. 62-00, C\'{o}digo Postal 76001, Santiago de Cali, Colombia}
\author{Jos\'{e} D. Ruiz-\'{A}lvarez}
\email{jose.ruiz@cern.ch}
\affiliation{Departamento de F\'{i}sica, Universidad de Los Andes, C\'{o}digo Postal 111711, Bogot\'{a}, Colombia}

\bigskip
\begin{abstract}	
In this work, the lepton-number-violating processes in $|\Delta L|=2$ decays of $\Lambda_b^0$ baryon, $\Lambda_b^0 \to p \pi^+\mu^- \mu^-$ and $\Lambda_b^0 \to \Lambda_c^+ \pi^+\mu^- \mu^-$, are investigated for the first time, via an intermediate on-shell Majorana neutrino $N$ with a mass in the GeV scale. We explore the experimental sensitivity of these  dimuon channels at the LHCb and CMS experiments, in which heavy neutrino lifetimes in the accessible ranges of $\tau_N = [1, 100, 1000]$ ps are considered. For a integrated luminosity collected of 10 and 50 fb${}^{-1}$ at the LHCb and 30, 300 and 3000 fb${}^{-1}$ at the CMS, we found significant sensitivity on branching fractions of the order ${\rm BR}(\Lambda_b^0 \to p \pi^+\mu^- \mu^-) \lesssim \mathcal{O}(10^{-9} - 10^{-8})$ and  ${\rm BR}(\Lambda_b^0 \to \Lambda_c^+ \pi^+\mu^- \mu^-) \lesssim \mathcal{O}(10^{-8} - 10^{-7})$. Exclusion regions on the parameter space ($m_N,|V_{\mu N}|^{2}$) associated with the heavy neutrino are presented and compared with those from $K^- \to \pi^+\mu^-\mu^-$ (NA48/2) and $B^- \to \pi^+\mu^-\mu^-$ (LHCb) as well as by different search strategies such as NA3, CHARMII, NuTeV, Belle, and DELPHI.
 \end{abstract} 


\maketitle
\bigskip

\section{Introduction}  \label{Intro}

At present, the possibility of extending the Standard Model (SM) by including right-handed sterile neutrinos with masses in the GeV-scale as an explanation of the neutrino mass generation, via a low-scale seesaw model, has taken strength both from the theoretical and experimental points of view~\cite{Rasmussen:2016}. On the theoretical side, seesaw scenarios with neutrinos masses below the electroweak scale are technically possible, without invoking higher energy scales~\cite{Rasmussen:2016,deGouvea:2007,Shaposhnikov:2005}. Additionally, these GeV-scale sterile neutrinos could also explain simultaneously the baryon asymmetry of the Universe via leptogenesis~\cite{Shaposhnikov:2005,Shaposhnikov:2013,Drewes:2014,GeV_Leptogenesis}. While on the experimental side, such sterile neutrinos may be produced and studied in a large variety of current and future experiments, both at the intensity and energy frontier, rendering it to a falsifiable scenario (for recent reviews, see Refs.~\cite{Deppisch:2015,Drewes:2013,Drewes:2015,deGouvea:2015,Fernandez-Martinez:2016} and references therein). 
 
 
An interesting search strategy for heavy Majorana neutrinos in the GeV range, it is to look for rare phenomena in which the total lepton number $L$ is broken by two units, generally referred to as $|\Delta L|=2$ processes. These sort of processes are forbidden in the SM and remain the best way to discern if neutrinos are Majorana fermions~\cite{deGouvea:2013}. The most appealing test of such a lepton-number-violating (LNV) processes is the neutrinoless double-$\beta$ ($0\nu\beta\beta$) decay~\cite{Rodejohann:2011,Gomez-Cadenas,Vissani:2015}. Although the case in which the exchange of a light massive Majorana neutrino is considered as the usual interpretation (standard mechanism)~\cite{Rodejohann:2011,Gomez-Cadenas,Vissani:2015}, recently,~Refs.~\cite{Drewes:2016lqo,Asaka:2016zib} have found that the rate for this process can also be enhanced due to a dominant contribution from heavy neutrino exchange with masses in the GeV-scale. Up to now, the $0\nu\beta\beta$ decay seems to be a rather elusive process and has not yet been observed experimentally. Currently, the best limits on their half-lives have been obtained from the nuclei $^{76}{\rm Ge}$ \cite{GERDA} and $^{136}{\rm Xe}$ \cite{EXO-200,KamLAND-Zen}. The non-observation allow us to set strong bounds on the mixing of a heavy neutrino $N$ with the electron ($V_{eN}$)~\cite{Kovalenko:2014}.

The low-energy studies of rare processes in $|\Delta L|= 2$ decays of pseudoscalar mesons and the $\tau$ lepton have been extensively studied~\cite{Atre:2009,Kovalenko:2000,Ali:2001,Atre:2005,Kovalenko:2005,Helo:2011,Cvetic:2010,
Zhang:2011,Bao:2013,Wang:2014,Quintero:2016,Sinha:2016,Gribanov:2001,Quintero:2011,Quintero:2012b,Quintero:2013,Quintero:2016,Sinha:2016,Dong:2013,Yuan:2013,
Quintero:2012a,Dib:2012,Dib:2014,Yuan:2017,Shuve:2016,Asaka:2016,Cvetic:2016,Zamora-Saa:2016},  as alternative LNV processes to $0\nu\beta\beta$ decay \cite{Atre:2009,Rodejohann:2011}. In these $|\Delta L|= 2$ decays, a sterile heavy neutrino with masses around 0.1 GeV $\lesssim m_N \lesssim$ few GeV can be produced on their mass-shell and its signal could be detected at different intensity frontier experiments. According to their final-state topology, they can be cla\-ssi\-fied as the following: 
\begin{itemize}
\item three-body channels \cite{Atre:2009,Kovalenko:2000,Ali:2001,Atre:2005,Kovalenko:2005,Helo:2011,Cvetic:2010,
Zhang:2011,Bao:2013,Wang:2014,Quintero:2016,Sinha:2016,Gribanov:2001,Dib:2014,Shuve:2016,Asaka:2016,Cvetic:2016,Zamora-Saa:2016} 
\begin{itemize}
\item $M^- \to M^{\prime +} \ell_i^- \ell_j^-$,   
\item $\tau^- \to \ell_\alpha^+ M^{\prime -} M^{\prime\prime -}$ ,
\end{itemize}
\item four-body channels \cite{Quintero:2011,Quintero:2012b,Quintero:2013,Quintero:2016,Sinha:2016,Dong:2013,Yuan:2013,
Quintero:2012a,Dib:2012,Yuan:2017,Cvetic:2016}
\begin{itemize}
\item $\bar{M}^{0} \to M^{\prime\prime +} M^{\prime +} \ell_\alpha^- \ell_\beta^-$ ,
\item $M^- \to M^{\prime\prime 0} M^{\prime +} \ell_i^- \ell_j^-$  ,
\item $\tau^{-} \to M^{\prime +} \nu_{\tau} \ell_\alpha^- \ell_{\beta}^-$,
\item $M^- \to \ell_i^- \ell_j^-  \ell^{\prime +} \nu_{\ell^\prime}$,
\end{itemize}
\end{itemize}

\noindent where $M \in \lbrace K, D, D_s, B, B_c \rbrace$ represents the decaying meson, $\ell_{i(j)}$ and $\ell^{(\prime)} \in \lbrace e, \mu, \tau \rbrace$ are the leptonic flavors, and $M^{\prime}$ and $M^{\prime\prime}$ represent final hadronic states that are allowed by kinematics. The possibility of $CP$ violation detection in $\Delta L=2$ decays of charged mesons \cite{Cvetic:CP} and the $\tau$ lepton \cite{Zamora-Saa:2016} have been also explored.


Experimentally, some of these $|\Delta L|= 2$ decays have been pursued for many years by different flavor facilities. No evidence has been seen so far, and u\-pp\-er limits on their branching fractions have been reported by the Particle Data Group (PDG) and several experiments such as NA48/2, BABAR, Belle, LHCb, and E791  \cite{PDG,CERNNA48/2:2016,BABAR,BABAR:2014,LHCb:2012,LHCb:2013,LHCb:2014,Belle:2011,Belle:2013,E791}. At CERN, further improvements are expected by the NA62 kaon factory \cite{NA62} and the LHCb in Run 2 and the future upgrade Run 3 \cite{LHCbUpgrade}. In addition, the forthcoming Belle II experiment aims to get $\sim$ 40 times more data than the those accumulated by its predecessor Belle (as well as BABAR) \cite{BelleII}. All these efforts will increase the sensitivity on $|\Delta L|= 2$ signals by 1 or 2 orders of magnitude. For instance, the future prospect of the Belle II search for the channel $B^- \rightarrow \pi^+\mu^- \mu^-$ has been discussed in Ref. \cite{Asaka:2016}. Furthermore, these improvements will allow the search for those signals not yet explored.

\begin{figure}[!t]
\centering
\includegraphics[scale=0.5]{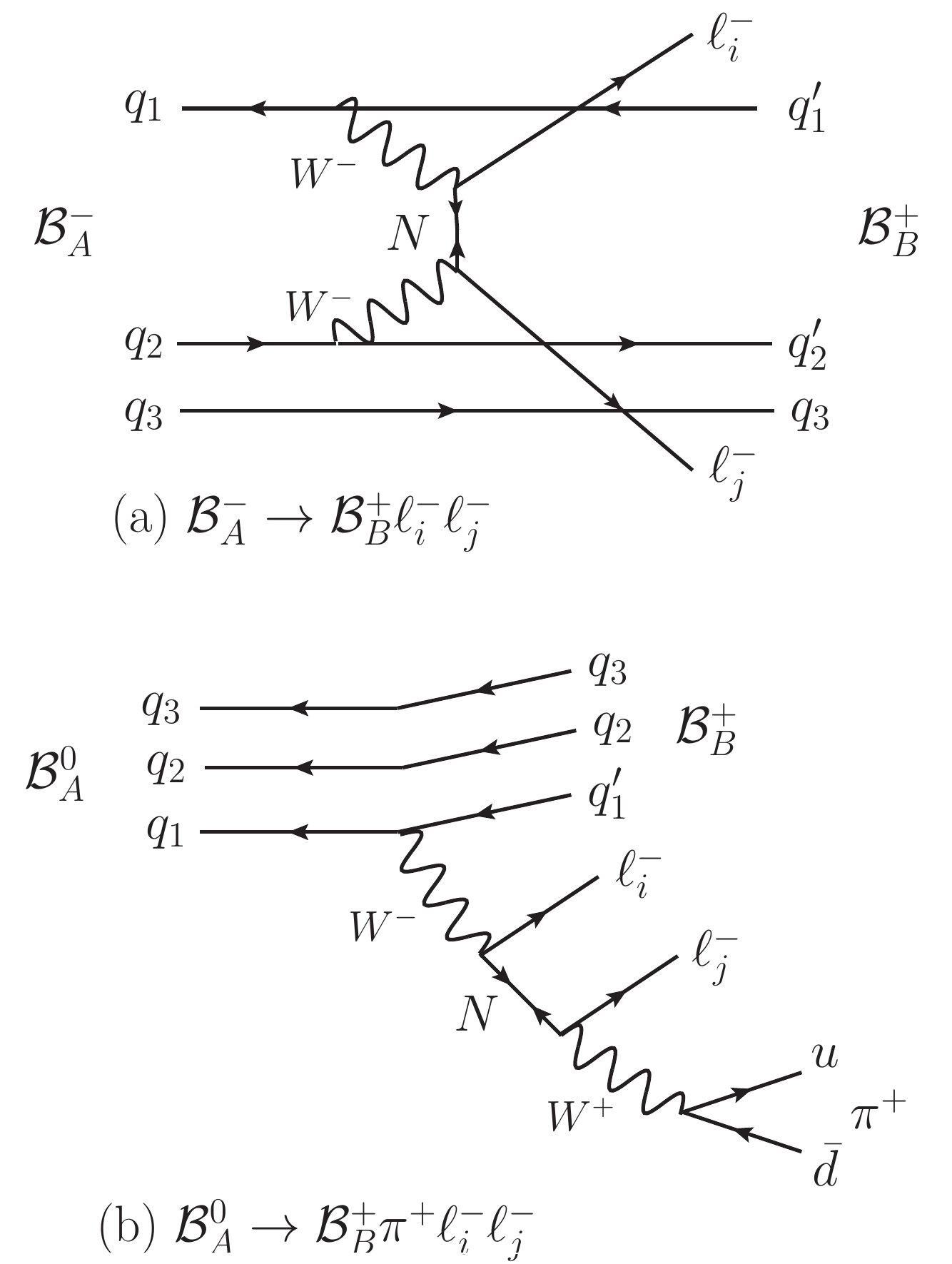}
\caption{\small Feynman graphs for $|\Delta L| = 2$ decays of the $\mathcal{B}_A$ baryon mediated by a Majorana neutrino $N$: (a) three-body $\mathcal{B}^- _A(q_1q_2q_3)\rightarrow \mathcal{B}_B^+(q_1^\prime q_2^\prime q_3)  \ell_i^- \ell_j^-$ and (b) four-body $\mathcal{B}^0 _A(q_1q_2q_3)\rightarrow \mathcal{B}_B^+(q_1^\prime q_2 q_3) \pi^+ \ell_i^- \ell_j^-$ channels. A similar diagram obtained from the lepton exchange $\ell_i \leftrightarrows \ell_j$ must be added.}
\label{Fig:1} 
\end{figure}

Aside from the LNV processes of pseudoscalar mesons and the $\tau$ lepton, the possibility of $|\Delta L|= 2$ decays of hyperons \cite{Shrock:1992,Lopez:2003,Lopez:2007} and charmed baryons \cite{Shrock:1992} has been also studied, namely, the three-body decay  $\mathcal{B}_A^- \rightarrow \mathcal{B}_B^+  \ell_i^- \ell_j^-$ as is graphically represented in Fig. \ref{Fig:1}(a), where $\mathcal{B}_A \ (\mathcal{B}_B)$ denotes an initial (final) baryon. From the experimental side, so far, the HyperCP \cite{HyperCP} and E653 \cite{E653:1995} Collaborations have reported limits on the branching fractions of $\Xi^- \rightarrow p \mu^-\mu^-$ and $\Lambda_c^+ \rightarrow \Sigma^- \mu^+\mu^+$, respectively.  With the large number of hyperons that are expected to be produced at the BESIII experiment, these $|\Delta L|= 2$ hyperon decays can be also searched \cite{BESIII:2016}. In addition to the $|\Delta L|= 2$ three-body decays of a charged baryon [Fig. \ref{Fig:1}(a)], $|\Delta L|= 2$ four-body decays of a neutral baryon $\mathcal{B}_A^0 \rightarrow \mathcal{B}_B^+ \pi^+  \ell_i^- \ell_j^-$ are possible as well, as is shown in Fig. \ref{Fig:1}(b).  In the case of the exchanged of a GeV-scale Majorana neutrino, this four-body process is generated through an $s$-channel in which the neutrino can be produced on-shell and dominates over the $t$-channel three-body one \cite{Kovalenko:2005}. 

Keeping this in mind and since the production of the $\Lambda_b^0$ baryon is around $\sim 5$\% of the total $b$-hadrons produced at the LHC, in this work, we will study new LNV processes in the four-body $|\Delta L|=2$ decays of $\Lambda_b^0$ baryon, $\Lambda_b^0 \to p \pi^+\mu^- \mu^-$ and $\Lambda_b^0 \to \Lambda_c^+ \pi^+\mu^- \mu^-$, in the scenario provided by the production of an on-shell Majorana neutrino. Within this simplified model approach, one heavy neutrino $N$ mixing with one flavor of SM lepton ($\ell=\mu$) and its interactions are completely determined by the mixing angle $V_{\mu N}$. Because of the relatively high muon reconstruction system, we focus on these same-sign dimuon channels and explored their expected sensitivity at the LHCb and CMS experiments. We will show that their experimental search allow us to put bounds on the parameter space associated with the mass $m_N$ and mixing $|V_{\mu N}|^2$ of the heavy Majorana neutrino. 

It is worth it to mention that the present work can be easily extendible to other $b-$baryons such as $\Omega_b, \Xi_b,$ and $\Sigma_b$, which are expected to be produced at the LHC in a lesser number than $\Lambda_b$.

This work is organized as follows. In Sec.~\ref{Sec-fourbodyleptonic}, we study the four-body $|\Delta L|=2$ decays of $\Lambda_b^0$ baryon. The expected experimental sensitivity for these channels at the LHCb and CMS experiments is presented in Sec. \ref{sensitivity}. In Sec.~\ref{constraints}, based on the results of the previous
sections, we estimate the constraints on the parameter space  $(m_N,|V_{\mu N}|^2)$ of the heavy neutrino that can be achieved, in which a comparison with different search strategies is also presented. Our conclusions are presented in Sec. \ref{Conclusion}.

\section{Four-body $|\Delta L|= 2$ decays of $\Lambda_b$ baryon}  \label{Sec-fourbodyleptonic}

In this section, we explore the four-body $|\Delta L| = 2$ decays of the $\Lambda_b^0$ baryon $\Lambda_b^0 \to \mathcal{B}^+ \pi^+\ell_i^- \ell_j^{-}$, which can occur via the exchange of a Majorana neutrino with a kinematically allowed mass, namely  $(m_{\ell_j} + m_\pi) < m_N < (m_{\Lambda_b} - m_{\mathcal{B}}-m_{\ell_i})$, where $\mathcal{B}^+ = p , \Lambda_c^+$ denotes a final-state baryon and $\ell_{i(j)}=e, \mu, \tau$. The corresponding diagram is shown in Fig. \ref{Fig:1}(b), with $ \mathcal{B}_A = \Lambda_b$ and $\mathcal{B}_B =  \mathcal{B}$. Among the possible same-sign dilepton final states, we will focus on the dimuon channels $\Lambda_b^0 \to (p, \Lambda_c^+) \pi^+\mu^- \mu^-$ and assume that only one on-shell heavy neutrino $N$ dominates these processes.

The  $|\Delta L| =2$ decays $\Lambda_b^0 \to (p, \Lambda_c^+) \pi^+\mu^- \mu^-$ receive the effect of a heavy Majorana neutrino with a mass in the ranges,
\bea
\Lambda_b^0 &\to & p \pi^+\mu^- \mu^- :  \ \ m_N \in [0.25,4.57] \ {\rm GeV}, \nonumber \\
\Lambda_b^0 &\to & \Lambda_c^+ \pi^+\mu^- \mu^- :  \ \ m_N \in [0.25,3.23] \ {\rm GeV}, \nonumber
\eea 

\noindent respectively. Within these mass ranges, the total decay width of the intermediate Majorana neutrino $N$ $(\Gamma_N)$ is much smaller than its mass, $\Gamma_N \ll m_N$ \cite{Atre:2009}, so the narrow width approximation is valid. This allows us to consider the Majorana neutrino as a particle that is produced on its mass shell through the semileptonic decay $\Lambda_b^0 \to \mathcal{B}^+ \mu^-N$, followed by the subsequent decay $N \to \mu^-\pi^+$. 

In this on-shell factorization approach, the branching fraction of $\Lambda_b^0 \to \mathcal{B}^+ \pi^+\mu^- \mu^-$ is then split into two subprocesses,
\bea \label{4leptonic}
{\rm BR}( \Lambda_b^0 \to \mathcal{B}^+\pi^+\mu^-\mu^-) &=& {\rm BR}(\Lambda_b^0 \to \mathcal{B}^+ \mu^- N) \nonumber\\
&& \times  \Gamma(N \to \mu^-\pi^+) \tau_N  /\hbar,
\eea 

\noindent  with $\tau_N$ as the lifetime of the Majorana neutrino. The decay width of $N \to \mu^-\pi^+$ is given by the expre\-ssi\-on \cite{Atre:2009}
\bea 
\Gamma(N & \to & \mu^-\pi^+) \nonumber \\
&=& \dfrac{G_F^2}{16 \pi}|V_{ud}^{\text{CKM}}|^2 |V_{\mu N}|^2 f_\pi^2 m_N \sqrt{\lambda(m_N^2,m_\mu^2,m_\pi^2)} \nonumber \\ 
&& \times \bigg[ \bigg(1- \dfrac{m_\mu^2}{m_N^2} \bigg)^2 - \dfrac{m_\pi^2}{m_N^2} \bigg(1+ \dfrac{m_\mu^2}{m_N^2} \bigg) \bigg], \label{Ntopimu}
\eea

\noindent where $G_F$ is the Fermi constant, $V_{ud}^{\text{CKM}}$ is the up-down Cabibbo-Kobayashi-Maskawa (CKM) matrix element, and $f_{\pi}$ is the pion decay constant. The usual kinematic K\"{a}llen function is denoted by $\lambda(x,y,z)=x^{2}+y^{2}+z^{2}-2(xy+xz+yz)$. The coupling of the heavy neutrino (sterile) $N$ to the charged current of lepton flavor $\mu$ is characterized by the quantity $V_{\mu N}$~\cite{Atre:2009}. Both its mass $m_N$ and $V_{\mu N}$ are unknown parameters that can be constrained (set) from the experimental non-observation (observation) of $|\Delta L| =2$ processes~\cite{Atre:2009,Helo:2011,Quintero:2016}. 

The lifetime of the Majorana neutrino $\tau_N=\hbar / \Gamma_N$ in \eqref{4leptonic} can be obtained by summing over all accessible final states that can be opened at the mass $m_N$~\cite{Atre:2009}. However, in further analysis (Secs. \ref{sensitivity} and \ref{constraints}), we will leave it as a phenomenological parameter accessible to the LHCb and CMS experiments.

To obtain the branching fraction of the semileptonic subprocess $\Lambda_b^0 \to \mathcal{B}^+  \mu^- N$, we begin from its amplitude, which is given by the expression 
\bea 
{\cal M}(& \Lambda_{b}^0 & \rightarrow \mathcal{B}^+\mu^-N)\nonumber \\ 
&=& \dfrac{G_F}{\sqrt{2}} V_{Qb}^{\text{CKM}} V_{\mu N} \langle \mathcal{B}(p_\mathcal{B}) |\bar{Q}\gamma_{\alpha}(1-\gamma_5) b|\Lambda_b (P) \rangle \nonumber \\ 
&& \times [\bar{u}(p_\mu) \gamma^{\alpha} (1-\gamma_{5}) v(p_N)], \label{AmpliA}
\eea

\noindent where $V_{Qb}^{\text{CKM}}$ is the CKM matrix element involved, with $Q = u, c$ for $\mathcal{B}=p, \Lambda_c$, respectively. The matrix element of the vector and axial-vector currents  associated to the baryonic transition $\Lambda_b \to \mathcal{B}$ can be parametrized as \cite{Detmold:2015} 
\bea
\left\langle \mathcal{B}(p_\mathcal{B}) |\bar{Q}\gamma_\alpha b|\Lambda_{b}(P)\right\rangle &= & \overline{u}_\mathcal{B}(p_\mathcal{B})\Big[ \gamma_{\alpha}f_{1}^V(t)+i \sigma_{\alpha\beta}q^{\beta}\frac{f_{2}^V(t) }{m_{\Lambda_b}} \nonumber \\
& & + q_{\alpha}\frac{f_{3}^V(t) }{m_{\Lambda_b}} \Big] u_{\Lambda_{b}}(P),
\eea

\bea
\left\langle \mathcal{B}(p_\mathcal{B}) |\bar{Q}\gamma_\alpha \gamma_5 b|\Lambda_{b}(P)\right\rangle &= & \overline{u}_\mathcal{B}	(p_\mathcal{B})\Big[ \gamma_{\alpha}f_{1}^A(t)+i \sigma_{\alpha\beta}q^{\beta} 	\frac{f_{2}^A(t) }{m_{\Lambda_b}} \nonumber \\
& & + q_{\alpha}\frac{f_{3}^A(t) }{m_{\Lambda_b}} \Big]  \gamma_5 u_{\Lambda_{b}}(P),
\eea 

\noindent in terms of six transition form factors $(f_{1}^V,f_{2}^V,f_{3}^V)$ and $(f_{1}^A,f_{2}^A,f_{3}^A)$, where $q=(P-p_\mathcal{B})$ is the transferred momentum and $t=q^2$. The spinors of $\mathcal{B}$ and $\Lambda_b$ are represented by $u_\mathcal{B}$ and $u_{\Lambda_b}$, respectively.



We end up with a branching ratio of $\Lambda_b^0 \to \mathcal{B}^+\mu^- N$ given by the following expression
\begin{widetext}
\bea \label{BR_Lambda_b}
{\rm BR}(& \Lambda_b^0 & \rightarrow \mathcal{B}^+ \mu^- N) \nonumber \\
&=& \dfrac{G_F^2 \tau_{\Lambda_b}}{512\pi^3 m_{\Lambda_b}^3 \hbar} |V_{Qb}^{\rm CKM}|^2 |V_{\mu N}|^2  \int_{(m_\mu +m_N)^2}^{\Delta_-^{2}} dt \ \sqrt{\lambda(m_\mu^2,m_N^2,t)\lambda(m_{\Lambda_b}^2,m_\mathcal{B}^2,t)}  \nonumber
\eea

\bea
&& \times \Big\lbrace \frac{16}{3t^3} [f_1^V(t)]^2\alpha_1^V(t) + \frac{8}{3 m_{\Lambda_b}^2 t^2} \big[f_2^V(t)\big]^2 \alpha_2^V(t) + \frac{8}{3 m_{\Lambda_b}^2 t} \big[f_3^V(t)\big]^2 \alpha_3^V(t) + \frac{32}{m_{\Lambda_b} t^2}\big[ f_1^V(t)f_2^V(t)\alpha_{12}^V(t)  \nonumber \\
&& \ \ \ + f_1^V(t)f_3^V(t)\alpha_{13}^V(t)  \big]+ \frac{16}{3t^3} [f_1^A(t)]^2\alpha_1^A(t) + \frac{8}{3 m_{\Lambda_b}^2 t^2} \big[f_2^A(t)\big]^2 \alpha_2^A(t) + \frac{8}{3 m_{\Lambda_b}^2 t} \big[f_3^A(t)\big]^2 \alpha_3^A(t) \nonumber \\
&& \ \ \ +\frac{32}{m_{\Lambda_b} t^2} \big[f_1^A(t)f_2^A(t)\alpha_{12}^A(t)  + f_1^A(t)f_3^A(t)\alpha_{13}^A(t)\big] \Big\rbrace ,
\eea

\noindent where the kinematic factors are

\bea
\alpha_1^{V/A}(t) &=& m_\mu^2 \big[ t (\Sigma_-^2 - 2 m_N^2\Sigma_+) - 2 t^2 (m_N^2 \mp m_{\Lambda_b} m_\mathcal{B} + \Delta_\mp^2) + 4m_N^2 \Sigma_-^2 + t^3\big] \nonumber \\
&& + (t-m_N^2) \big[ m_N^2 (t \Sigma_+ -2\Sigma_- +t^2) - t(\Delta_\mp^2 - t)(\Sigma_\pm^2 +2t)\big] , \\
\alpha_2^{V/A}(t) &=& \big[t (m_\mu^2 + m_N^2) + (m_\mu^2 - m_N^2)^2 - 2t^2 \big] \big[t (\Sigma_\pm^2 \pm 4m_{\Lambda_b} m_\mathcal{B}) - 2 (\Sigma_-^2 + t^2)\big] ,\\
\alpha_3^{V/A}(t) &=&  (\Delta_{\pm}^2 - t^2) \big[m_\mu^2(t + 2m_N^2 -m_\mu^2) + m_N^2 (t- m_N^2) \big] ,\\
\alpha_{12}^{V /A}(t) &=&  \Delta_{\pm} (\Delta_{\mp} - t) \big[m_\mu^2(t - 2m_N^2) +m_\mu^4 + m_N^4 +m_N^2 t -2t^2\big],\\
\alpha_{13}^{V/A}(t) &=&\Delta_{\mp} (\Delta_{\pm} - t) \big[ (m_\mu^2 - m_N^2)^2 -t(m_\mu^2 + m_N^2)\big],
\eea
\end{widetext}

\noindent with $\Delta_{\pm}= m_{\Lambda_b} \pm m_\mathcal{B}$ and $\Sigma_{\pm}= m_{\Lambda_b}^2 \pm m_\mathcal{B}^2$. As a cross-check, we have verified that this expression is consistent with the one obtained in Ref.~\cite{Ramazanov:2008ph}. For ensuing numerical evaluations in Sec.~\ref{constraints}, we will use the theoretical predictions obtained by Lattice QCD on the form factors $(f_{1}^V,f_{2}^V,f_{3}^V)$ and $(f_{1}^A,f_{2}^A,f_{3}^A)$~\cite{Detmold:2015}. Besides, we will take the following numerical inputs: $|V_{ud}^{\text{CKM}}|= 0.97417$, $|V_{cb}^{\rm{CKM}}| = 40.5 \times 10^{-3}$, $|V_{ub}^{\text{CKM}}|= 4.09 \times 10^{-3}$~\cite{PDG}, and $f_{\pi}= 130.2(1.7)$ MeV~\cite{Rosner:2015}. The masses of the particles involved and lifetime $\tau_{\Lambda_b}$ are taken from Ref.~\cite{PDG}.

We close by mentioning that there are different calculations of the $\Lambda_b \to (p, \Lambda_c)$ form factors in the literature, for instance, the covariant confined quark model~\cite{Gutsche:2015mxa}. Using this model, we have checked that one gets very similar results as the ones presented in Sec.~\ref{constraints} by means of Lattice QCD~\cite{Detmold:2015}.

\section{Expected experimental sensitivity at the LHC}  \label{sensitivity}

In this section, we provide an estimation of the expected number of events at the LHC, namely, LHCb and CMS experiments, for the $|\Delta L|=2$ channels $\Lambda_{b}^0\to \mathcal{B}^+\pi^+\mu^-\mu^-$ (with $\mathcal{B} = p, \Lambda_c$), discussed above.

\subsection{LHCb experiment} \label{LHCb}

The number of expected events in the LHCb experiment has the form 
\bea
\label{N:LHCb}
N_{\rm exp}^{\rm LHCb} &=&  \sigma(pp\to H_b X)_{\rm acc}f(b\to \Lambda_b) {\rm BR}(\Lambda_{b}^0 \to \Delta L=2)\nonumber \\
&& \times  \epsilon_D^{\rm LHCb}(\Lambda_{b}^0 \to \Delta L=2) P_N^{\rm LHCb} \ \mathcal{L}^{\rm LHCb}_{\rm int},
\eea

\noindent where $\sigma(pp\to H_b X)_{\rm acc}$ is the production cross-section of $b$-hadrons inside the LHCb geometrical acceptance; $f(b\to \Lambda_b)$ is the hadronization factor of a $b$-quark to $\Lambda_b^0$ baryons; $\mathcal{L}_{\rm int}^{\rm LHCb}$ is the integrated luminosity; ${\rm BR}(\Lambda_b^0\to \Delta L=2)$ corresponds to the branching fraction of the given LNV process; and $\epsilon_D^{\rm LHCb}(\Lambda_{b}^0 \to \Delta L=2)$ is its detection efficiency of the LHCb detector involving  reconstruction, selection, trigger, particle misidentification, and detection efficiencies. Most of the the on-shell neutrinos produced in the decays  $\Lambda_{b}^0\to (p,\Lambda_{c}^+)\mu^-N$ are expected to live a long enough time to travel through the detector and decay ($N \to  \pi^+\mu^-$) far from the interaction region. This effect is given by the $P_N^{\rm LHCb}$ factor (acceptance factor), which accounts for the probability of the on-shell neutrino $N$ decay products to be inside the LHCb detector acceptance~\cite{Dib:2014}. The reconstruction efficiency will depend on this acceptance factor as well.

The production cross section is well measured to be $\sigma(pp\to H_b X)_{\rm acc}=(75.3 \pm 5.4 \pm  13.0)$~$\mu$b inside the LHCb acceptance \cite{Aaij:2010gn}. The hadronization factor can be related with the total hadronization factor to baryons as   $f(b\to \text{baryons})\simeq f(b\to\Lambda_b)(1 + 2 f(b\to B_s^0)/f(b\to B^0))$, where we have used isospin symmetry described in Ref.~\cite{Isos:isospinarguments}. Thus, the hadronization factor can be built from Ref~\cite{HFAG}, where  $f(b\to \text{baryons}) = 0.088 \pm 0.012$, $f(b\to B_s^0) = 0.103 \pm 0.005$ and $f(b\to B^0) = 0.404\pm 0.006$, where these factors are computed as an average of LEP and Tevatron measurements. This leads to $f(b\to\Lambda_b)\simeq 0.053 \pm 0.017$.


\begin{figure}[!b]
\centering
\includegraphics[scale=0.47]{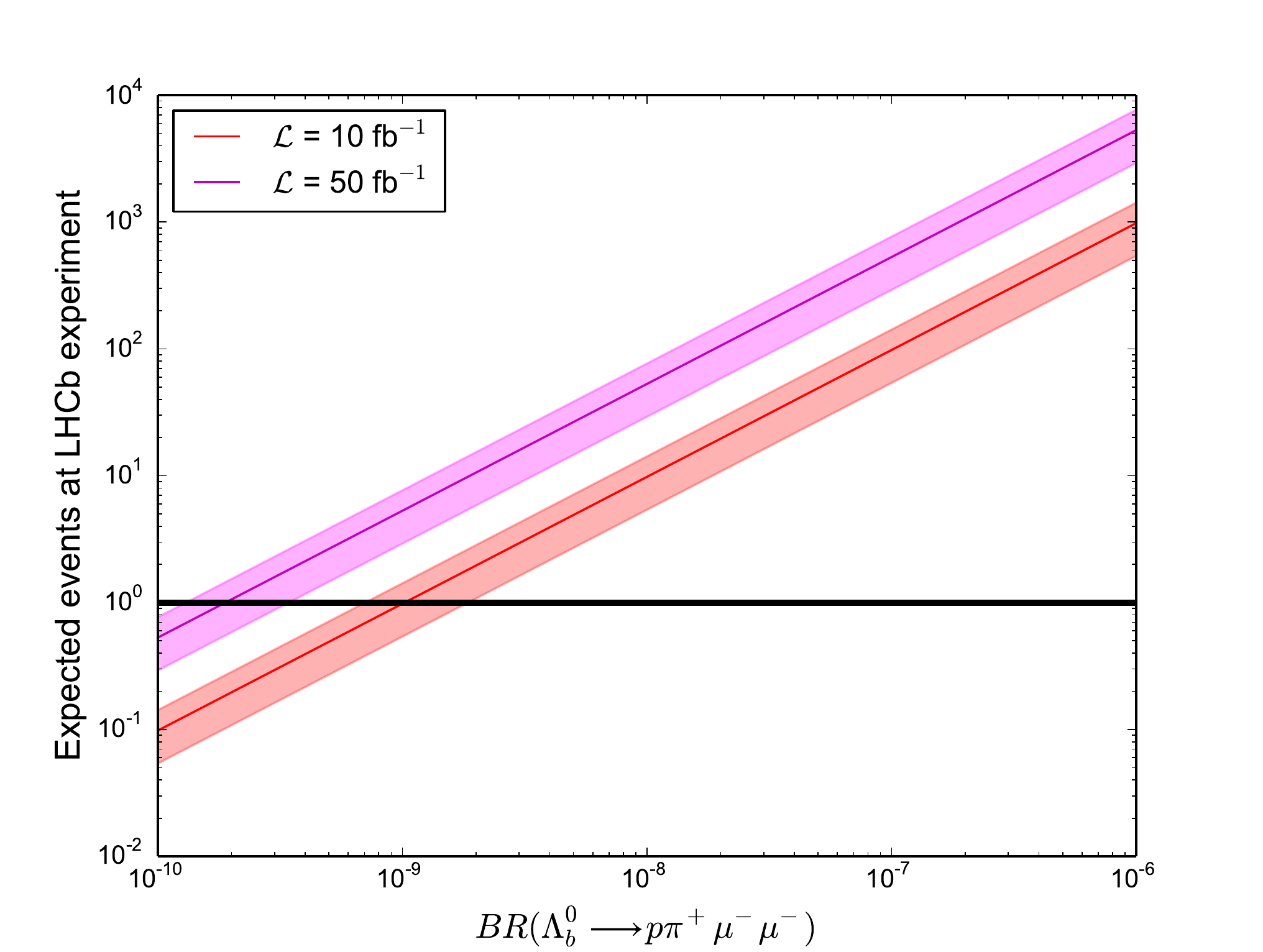} 
\includegraphics[scale=0.47]{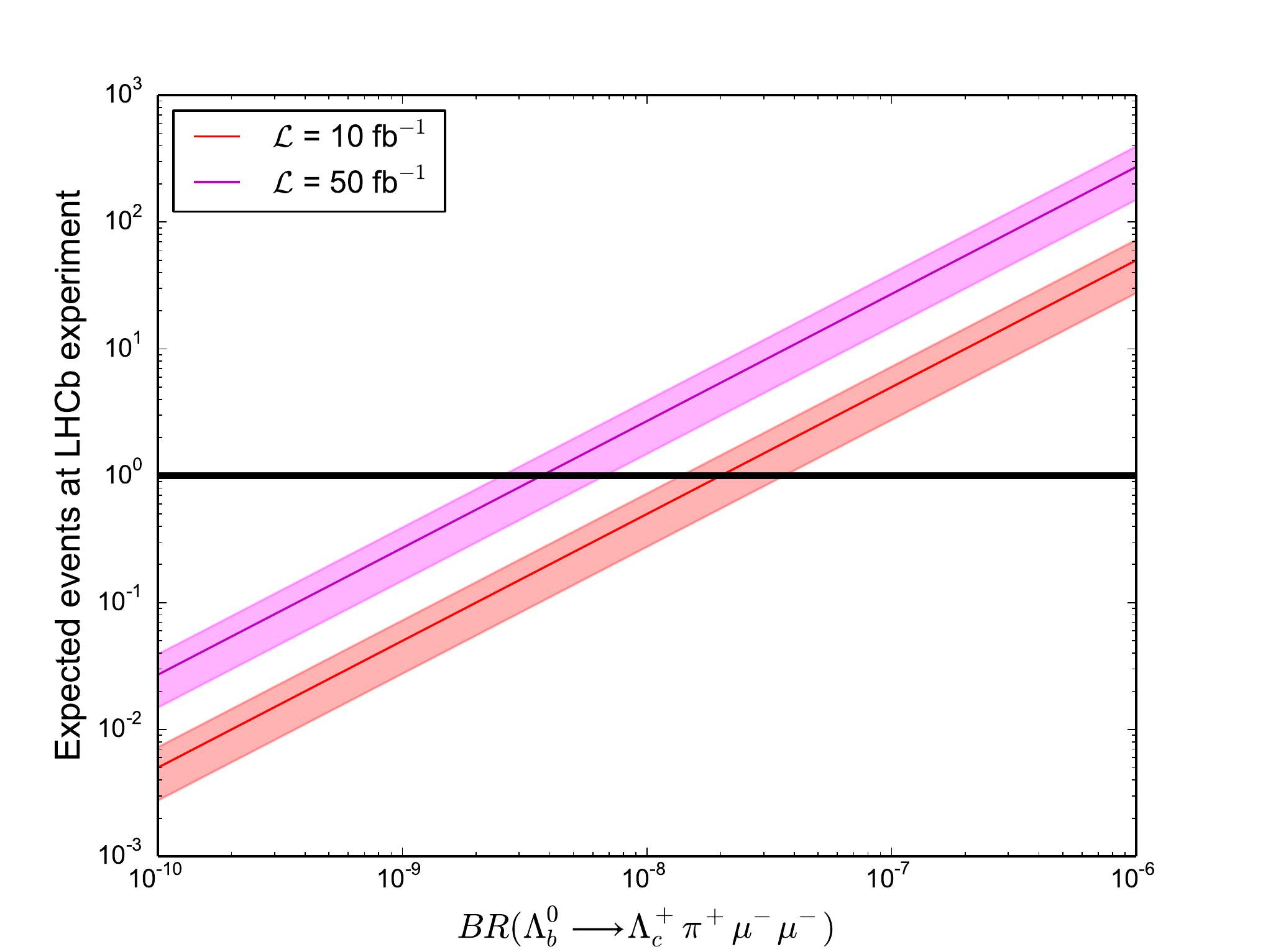}
\caption{\small  Number of expected events of the process $\Lambda_{b}^0\to p\pi^+\mu^-\mu^-$ (top) and $\Lambda_{b}^0\to \Lambda_c^+\pi^+\mu^-\mu^-$ (bottom) to be observed in the LHCb experiment as a function of their branching fractions for a luminosity of 10 fb$^{-1}$ (red) and 50 fb$^{-1}$ (magenta). Solid lines show the central value, while the filled area shows the 1-$\sigma$ uncertainty.}
\label{Fig:2} 
\end{figure}

Precise computation of the detection efficiency requires fully simulated decay-specific Monte Carlo samples, reconstructed in the same manner as real data and with a simulation of the full detector.  However a rough estimation can be done with detection efficiencies already reported by the LHCb experiment in the study of some $\Lambda_b^0$ decays with the same or similar final-state particle content as our LNV modes, such as Ref.~\cite{Aaij:2017awb}. Here, the study of $\Lambda_b^0 \to  [c\bar c]pK−$ modes ( $[c\bar c]$ stands for $J\psi$, $\chi_{c1}$, and $\chi_{c2}$ charmonia states which decay into a pair of muons tracks) is performed, and the measured yields and efficiency corrected yields are given, and therefore $\epsilon_D^{\rm LHCb}$ per mode can be extracted. From that information, it is extracted that $\epsilon( \Lambda_b^0 \to  J/\psi(\to\mu^+\mu^-)pK−)\simeq 0.0246\pm0.0001$, where it  must be mentioned that tight selection criteria are used, to maximize signal over background, given the large suppression of the decays under study, something similar to the $|\Delta L| = 2$ decays of interest. On the other hand, in Ref.~\cite{Aaij:2017ewm},  it is determined the following ratio of efficiencies $\epsilon( \Lambda_b^0 \to \mu^+\mu^-p\pi−)/\epsilon( \Lambda_b^0 \to  J/\psi(\to\mu^+\mu^-)p\pi−) =0.49\pm 0.02$, which are channels with identical topologies to $\Lambda_b\to p\pi^+\mu^-\mu^-$, therefore we can safely state $\epsilon_D^{\rm LHCb}(\Lambda_b^0\to p\pi^+\mu^-\mu^-)\simeq 0.0121\pm 0.0005$. As, $ \Lambda_b^0 \to \Lambda_c^+(\to pK^-\pi^+)\pi^+\mu^-\mu^-$ involves two additional charged tracks, we can multiply above expression for 90\% for each additional track, keeping same uncertainty in a conservative scenario, to obtain $\epsilon_D^{\rm LHCb}(\Lambda_b^0\to \Lambda_c^+\pi^+\mu^-\mu^-)\simeq 0.0098\pm0.0005$. Finally, in Ref.~\cite{Aaij:2016xmb}, reconstruction efficiencies for hypothetical long-lived particles inside the LHCb acceptance are given. Here we can observe that a maximum variation of about 25\% is measured in the efficiencies of particles living in the [5 - 100]~ps range, with masses up to 200~GeV/$c^{2}$. Thus, to account for this effect, we will just add a 25\% relative uncertainty to our efficiency prediction, obtaining finally
\bea
\epsilon_D^{\rm LHCb}(\Lambda_b^0\to p\pi^+\mu^-\mu^-) P_N^{\rm LHCb} &\simeq & 0.0121\pm 0.0030, \nonumber \\ \nonumber
\epsilon_D^{\rm LHCb}(\Lambda_b^0\to \Lambda_c^+\pi^+\mu^-\mu^-)P_N^{\rm LHCb}   &\simeq & 0.0098\pm 0.0025 .\nonumber
\eea

\noindent The combination of all inputs to the number of expected events leads to a relative uncertainty of 45\% in both LNV modes, where to compute $N(\Lambda_b^0 \to \Lambda_c^+(\to pK^-\pi^+)\pi^+\mu^-\mu^-)$ we have used  ${\rm BR}(\Lambda_c^+ \to pK^-\pi^+)=(6.35\pm0.33)\%$.

Assuming above assumptions on the efficiency and cross section, in Figs.~\ref{Fig:2}[top] and~\ref{Fig:2}[bottom] we plot the number of expected events to be observed in the LHCb experiment as a function of the branching fraction for $\Lambda_{b}^0\to p\pi^+\mu^-\mu^-$ and $\Lambda_{b}^0\to \Lambda_c^+\pi^+\mu^-\mu^-$, respectively. The red and magenta regions correspond to a luminosity of $\mathcal{L}_{\rm int}^{\rm LHCb} = 10$ and $50 \ {\rm fb}^{-1}$ for LHC Run2 and LHC Run3, respectively. In Table~\ref{BR:LHCb} we present the number of expected events for some selected values of branching ratios. We can see that values of branching fractions of the order $10^{-9} - 10^{-8}$ for $\Lambda_{b}^0\to p\pi^+\mu^-\mu^-$ and $10^{-8} - 10^{-7}$ for $\Lambda_{b}^0\to \Lambda_c^+\pi^+\mu^-\mu^-$ might be within the experimental reach of the LHCb.
 
\begin{table}[!t]
\centering
\renewcommand{\arraystretch}{1.2}
\renewcommand{\arrayrulewidth}{0.8pt}
\caption{\small Number of expected events at the LHCb for some selected values of the branching ratio of $\Lambda_{b}^0\to p\pi^+\mu^-\mu^-$ and $\Lambda_{b}^0\to \Lambda_c^+\pi^+\mu^-\mu^-$.}
\begin{tabular}{cccc}
\hline\hline \rowcolor{Gray}
Mode & $\mathcal{L}_{\rm int}^{\rm LHCb}$ (fb$^{-1}$) & BR & Number of events \\
\hline
 &  & $10^{-6}$ & $981 \pm 441$\\
 &  10 & $10^{-7}$ & $98 \pm 44$\\
 &   & $10^{-8}$ & $10 \pm 4$\\
\cline{2-4}
$\Lambda_{b}^0\to p\pi^+\mu^-\mu^-$ & & $10^{-7}$ & $530 \pm 238$ \\ 
  & 50  & $10^{-8}$ & $53 \pm 24$ \\ 
    &  & $10^{-9}$ & $5 \pm 2$ \\ 
 \hline
 & 10  & $10^{-6}$  & $50 \pm 23$\\
 &   & $10^{-7}$  & $5 \pm 2$\\
  \cline{2-4}
  $\Lambda_{b}^0\to \Lambda_c^+\pi^+\mu^-\mu^-$   &  & $10^{-6}$ & $272 \pm 122$\\
 & 50  & $10^{-7}$ & $27 \pm 12$ \\ 
  &  & $10^{-8}$ & $3 \pm 1$ \\ 
\hline\hline
\end{tabular} \label{BR:LHCb}
\end{table}

\subsection{CMS experiment} \label{CMS}

For the CMS experiment, the number of expected events is given by the expression
\bea
\label{N:CMS}
N_{\rm exp}^{\rm CMS} &=&   \sigma(pp\to\Lambda_{b}^0 X)  {\rm BR}(\Lambda_{b}^0 \to \Delta L=2) \nonumber \\
&& \times \epsilon_D^{\rm CMS}(\Lambda_{b}^0 \to \Delta L=2) P_N^{\rm CMS} \mathcal{L}_{\rm int}^{\rm CMS},
\eea

\noindent where $\mathcal{L}_{\rm int}^{\rm CMS}$ is the integrated luminosity, $\sigma(pp\to\Lambda_{b}^0 X)$ is the production cross-section of $\Lambda_{b}^0$ baryons inside the CMS geometrical acceptance proton-proton collisions, $P_N^{\rm CMS}$ is the acceptance factor at the CMS, $\epsilon_D^{\rm CMS}(\Lambda_{b}^0 \to \Delta L=2)$ is the CMS experiment efficiency that involves the detection and trigger efficiencies and the geometrical acceptance to detect our signal, and BR$(\Lambda_{b}^0 \to \Delta L=2)$ is its respective branching ratio.

To calculate the efficiency of the CMS experiment to accept our signal we have to take into account the track reconstruction efficiency for charged pions (due to the lack of the particle identification, CMS assumes that all charged tracks are pions) as well as the muon reconstruction efficiency for the kinematics of the signal. The produced particles from $\Lambda_{b}^0$ decay have relatively low $p_{T}$. We consider that the pions and muons from our signal have mainly a $p_{T}<20$ GeV. From~\cite{Chatrchyan:2014fea} CMS reconstruction efficiency of charged tracks in the tracker, for the $p_{T}$ spectrum of interest, the reconstruction efficiency of pions from our signal is of 90\% at 7 TeV proton-proton collisions. We assume that this efficiency remains mainly unchanged at 13 TeV. At 8 TeV the muon reconstruction efficiency for a $p_{T}>3$ GeV and $p_{T}<20$ GeV has been measured to be around 90\%~\cite{MuonEff}. We also assume this efficiency remains unchanged at 13 TeV. 


Additionally, precise computation of the detection efficiency of our signal events requires fully simulated decay specific Monte Carlo samples, reconstructed using the same techniques as in real data and with a full simulation of the detector. However, a rough estimation can be done with detection efficiency already reported by the CMS experiment in the study of the $\Lambda_b^0$ baryon cross section~\cite{cms:lambdabcrosssection}, which is 0.73\% (with an uncertainty of approximately 10\%). For the decay channel $\Lambda_{b}^0 \to p\pi^+\mu^-\mu^-$ we have the same final-state particle, and thus we assume the same reconstruction efficiency, i.e., $\epsilon_D^{\rm CMS}(\Lambda_{b}^0 \to p\pi^+\mu^-\mu^-) = (0.73 \pm 0.07)\%$. For the channel $\Lambda_{b}^0 \to \Lambda_c^{+}\pi^+\mu^-\mu^-$, there are two other tracks (considering the chain decay $\Lambda_c^{+} \to pK^-\pi^+$). For this reason, the efficiency detection will be reduced, and taking into account the efficiency of these additional tracks, we will assume it to be $\epsilon_D^{\rm CMS}(\Lambda_{b}^0 \to \Lambda_c^+\pi^+\mu^-\mu^-) = (0.59 \pm 0.06)\%$. This is not an optimistic case, since the CMS Collaboration is making an important effort to improve the reconstruction capabilities in Run 2. 


We have considered a minimum and maximum neutrino lifetime of $\tau_N= 1$ ps and 1000 ps, respectively, where the detector has sensitivity. Considering that the neutrino travels at nearly the speed of light and taking into account that the neutrino comes from the $\Lambda_{b}^0$ decay, the decay length of the neutrino is $L_N= 0.3$ cm (30 cm) for $\tau_N= 1$ ps (1000 ps) lifetime.  We consider that the decay length of $\Lambda_{b}^0$ is 0.4 cm because its lifetime is 1.466 ps~\cite{PDG}. According to Ref.~\cite{Chatrchyan:2014fea}, the reconstruction efficiency for tracks originated at a distance of 30 cm from the collision point is 55\% and for 1 cm is 100\%, where we can observe a maximum variation of about 18\%. Then, to consider this effect, we will add an 18\% relative uncertainty to our efficiency prediction. Therefore, we estimate an acceptance factor  $0.55 \leq P_N \leq 1$ for  1 ps $\leq \tau_N \leq$ 1000 ps. Of course, this is an estimation from studies performed by the CMS experiment at 7 TeV, and a precise estimation will require knowledge of the production and decay vertex, which can be adequately included during the data analysis. Finally, the efficiencies will be
\bea
\epsilon_D^{\rm CMS} (\Lambda_{b}^0 \to p\pi^+\mu^-\mu^-) P_N^{\rm CMS} &\simeq & 0.073 \pm 0.015, \nonumber \\ \nonumber
\epsilon_D^{\rm CMS} (\Lambda_{b}^0 \to \Lambda_c^{+}\pi^+\mu^-\mu^-) P_N^{\rm CMS} &\simeq & 0.059 \pm 0.013 .\nonumber
\eea

The cross section times the branching fraction $\sigma(\Lambda_{b})  \times{\rm BR}(\Lambda_{b}\to J/\psi  \Lambda)$ [with $p_T(\Lambda_b) > 10$ GeV and $| y(\Lambda_b)|<2.0$] at 7 TeV measured by the CMS experiment is $1.16\pm 0.06\pm 0.12$ nb~\cite{cms:lambdabcrosssection}. It is not possible to directly infer the BR$(\Lambda_{b}\to J/\psi  \Lambda)$ because there is not a published measurement of $f(b \to \Lambda_{b})$. However, we will use the value estimated in the previous section $f(b\to\Lambda_b)\simeq 0.053 \pm 0.017$. Now, the PDG reports BR$(\Lambda_{b}\to J/\psi  \Lambda) \times f(b \to \Lambda_{b}) = (5.8 \pm 0.8) \times 10^{-5}$. Using the previous numbers, we can find a cross section value of $\sigma(pp\to\Lambda_{b} X)=(1.06 \pm 0.39)$ $\mu$b at 7 TeV . Extrapolating this value to 13 TeV assuming that the cross sections grows as the energy collision $\sigma(pp\to\Lambda_{b} X)=(1.97 \pm 0.72)$ $\mu$b. 

\begin{figure}[!t]
\centering
\includegraphics[scale=0.44]{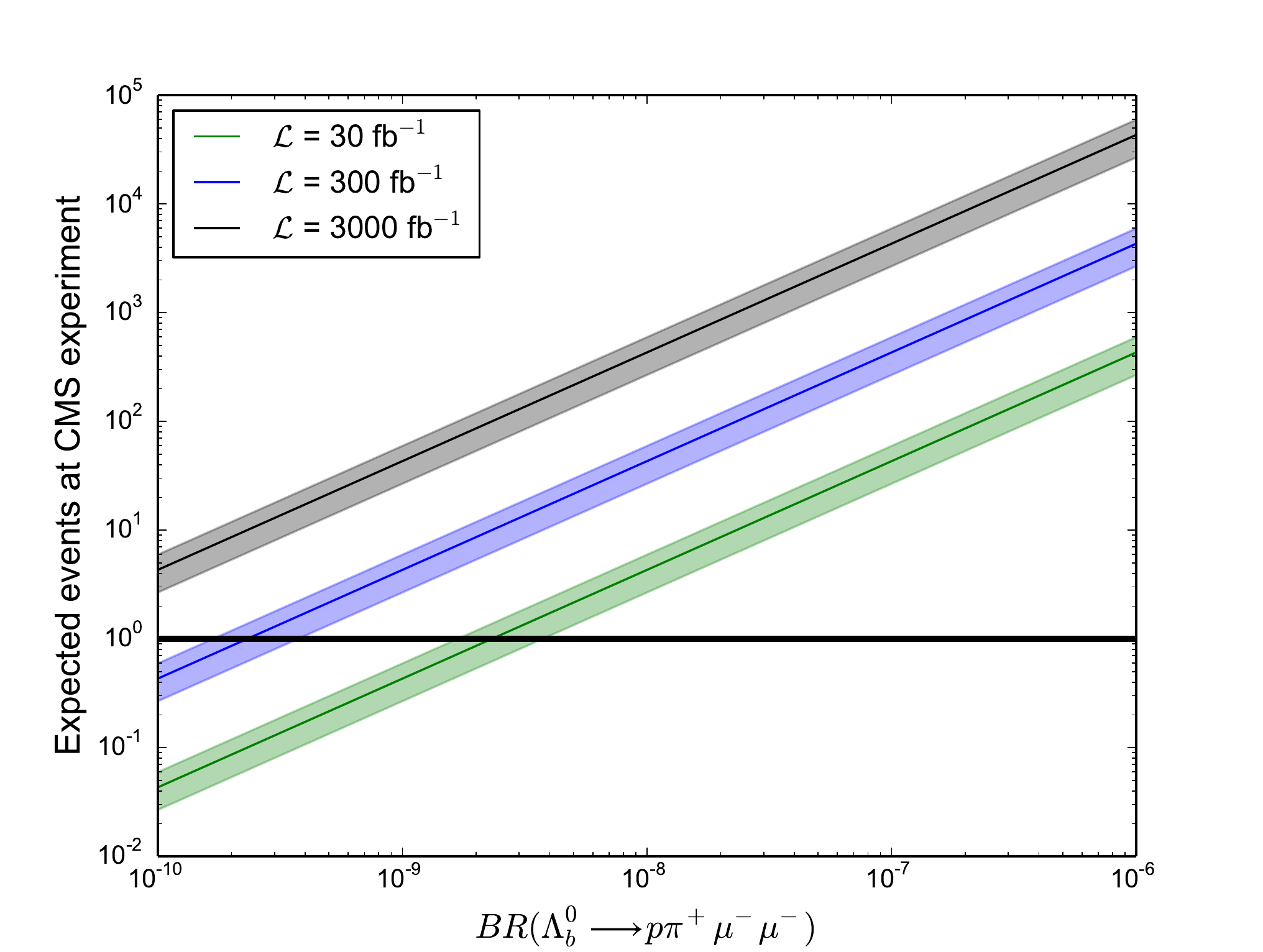} 
\includegraphics[scale=0.44]{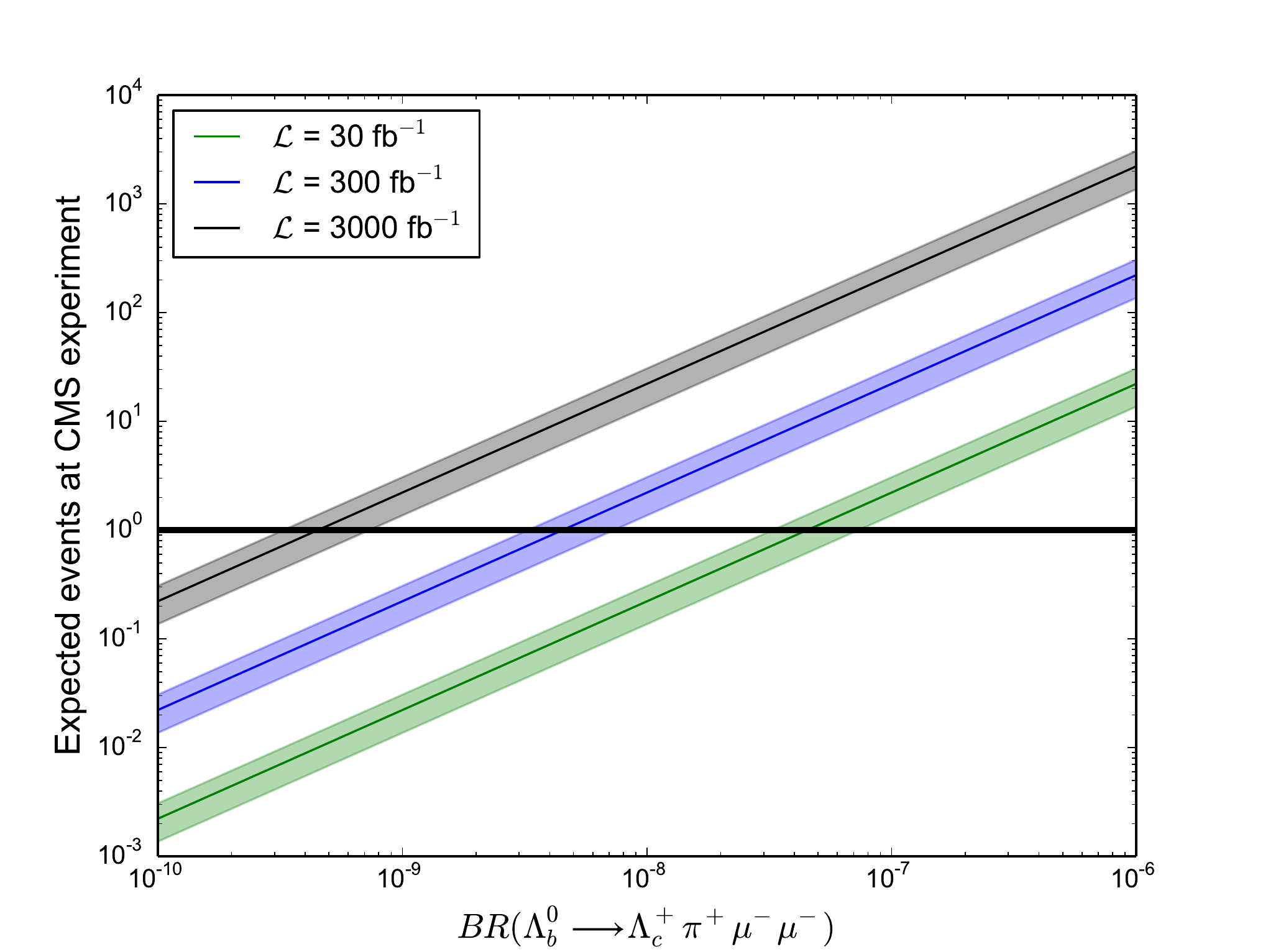}
\caption{\small  Number of expected events of the process $\Lambda_{b}^0\to p\pi^+\mu^-\mu^-$ (top) and $\Lambda_{b}^0\to \Lambda_c^+\pi^+\mu^-\mu^-$ (bottom])to be observed in the CMS experiment as a function of their branching fractions for a luminosity of 30 fb$^{-1}$ (green), 300 fb$^{-1}$ (blue), and 3000 fb$^{-1}$ (gray). Solid lines shows the central value, while the filled area shows the 1-$\sigma$ uncertainty.}
\label{Fig:3} 
\end{figure}

Considering the values found for the cross section and efficiencies, Fig.~\ref{Fig:3} shows the number of expected events to be observed in the CMS experiment for $\mathcal{L}_{\rm int}^{\rm CMS}=$ 30, 300, and 3000 fb$^{-1}$ as a function of the branching ratio of $\Lambda_{b}^0$ into our LNV channels, where to compute $N(\Lambda_b^0 \to \Lambda_c^+(\to pK^-\pi^+)\pi^+\mu^-\mu^-)$ we have used  ${\rm BR}(\Lambda_c^+ \to pK^-\pi^+)=(6.35\pm0.33)\%$. For integrated luminosities of 30 and 300 fb$^{-1}$, in Table~\ref{BR:CMS}, we present the number of expected events for some selected values of branching ratios. It is easy to see that for 3000 fb$^{-1}$ the number of events will increase 1 order of magnitude, since these scale in the same way as the luminosity. 
We found significant sensitivity at the CMS on branching fractions of the order $10^{-9} - 10^{-8}$ for $\Lambda_{b}^0\to p\pi^+\mu^-\mu^-$ and  $10^{-8} - 10^{-7}$ for $\Lambda_{b}^0\to \Lambda_c^+\pi^+\mu^-\mu^-$.

\begin{table}[!t]
\centering
\renewcommand{\arraystretch}{1.2}
\renewcommand{\arrayrulewidth}{0.8pt}
\caption{\small Number of expected events at the CMS for some selected values of the branching ratio of $\Lambda_{b}^0\to p\pi^+\mu^-\mu^-$ and $\Lambda_{b}^0\to \Lambda_c^+\pi^+\mu^-\mu^-$.}
\begin{tabular}{cccc}
\hline\hline \rowcolor{Gray}
Mode & $\mathcal{L}_{\rm int}^{\rm CMS}$ (fb$^{-1}$) & BR & Number of events \\
\hline
 &  & $10^{-6}$ & $431 \pm 164$\\
 &  30 & $10^{-7}$ & $43 \pm 16$\\
 $\Lambda_{b}^0\to p\pi^+\mu^-\mu^-$ &   & $10^{-8}$ & $4 \pm 2$\\
\cline{2-4}
 & 300 & $10^{-8}$ & $43 \pm 16$ \\ 
  &  & $10^{-9}$ & $4 \pm 2$ \\ 
 \hline
 & 30  & $10^{-6}$  & $27 \pm 10$\\
  $\Lambda_{b}^0\to \Lambda_c^+\pi^+\mu^-\mu^-$ &   & $10^{-7}$  & $3 \pm 1$\\
  \cline{2-4}
   &  300  & $10^{-7}$ & $27 \pm 10$ \\ 
  &  & $10^{-8}$ & $3 \pm 1$ \\ 
\hline\hline
\end{tabular} \label{BR:CMS}
\end{table}

In the analysis of the next section, we will take these values of branching fractions as the most conservative ones and accessible to the LHCb and CMS experiments.

\section{Constraints on the $(m_N,|V_{\mu N}|^2)$ plane}  \label{constraints}

Experimental limits from the search of $|\Delta L| =2$ processes can be  reinterpreted as constraints on the parameter space of a heavy sterile neutrino $(m_N,|V_{\mu N}|^2)$, namely, the squared mixing element $|V_{\mu N}|^2$ as a function of the mass  $m_N$~\cite{Atre:2009,Helo:2011,Quintero:2016}. Based on the analysis presented in the previous section, here we explore the constraints on the $(m_N,|V_{\mu N}|^2)$ plane that can be achieved from the experimental searches on $\Lambda_b^0 \to  (p, \Lambda_c^+) \pi^+\mu^- \mu^-$ at the LHC. 

From Eq.~\eqref{4leptonic}, it is straightforward to obtain the relation
\beq
|V_{\mu N}|^2 = \Bigg[  \frac{\hbar \ {\rm BR}(\Lambda_b^0 \to  \mathcal{B}^+ \pi^+\mu^- \mu^-)}{ \overbar{\rm BR}(\Lambda_b^0 \to \mathcal{B}^+ \mu^- N) \times \overbar{\Gamma}(N \to \mu^-\pi^+) \tau_N}\Bigg]^{1/2}
\eeq


\noindent where
\bea
\overbar{\rm BR}(\Lambda_b^0 \to \mathcal{B}^+ \mu^- N) &=& {\rm BR}(\Lambda_b^0 \to \mathcal{B}^+ \mu^- N) / |V_{\mu N}|^2 , \nonumber \\
&& \\
\overbar{\Gamma}(N \to \mu^-\pi^+) &=& \Gamma(N \to \mu^-\pi^+)/ |V_{\mu N}|^2,
\eea

\noindent are the normalized branching ratio ${\rm BR}(\Lambda_b^0 \to  \mathcal{B}^+ \mu^- N)$ [Eq. \eqref{BR_Lambda_b}] and decay width $\Gamma(N \to \mu^-\pi^+)$ [Eq. \eqref{Ntopimu}] to the neutrino mixing $|V_{\mu N}|^2$. As was already discussed in Sec. \ref{sensitivity} and following the analysis of NA48/2~\cite{CERNNA48/2:2016} and the LHCb~\cite{BABAR:2014}, we will consider heavy neutrino lifetimes of $\tau_N = [1, 100, 1000]$ ps as benchmark points in our analysis. This will allow us to extract limits on $|V_{\mu N}|^2$ without any additional assumption on the relative size of the mixing matrix elements.

\begin{figure}[!t]
\centering
\includegraphics[scale=0.44]{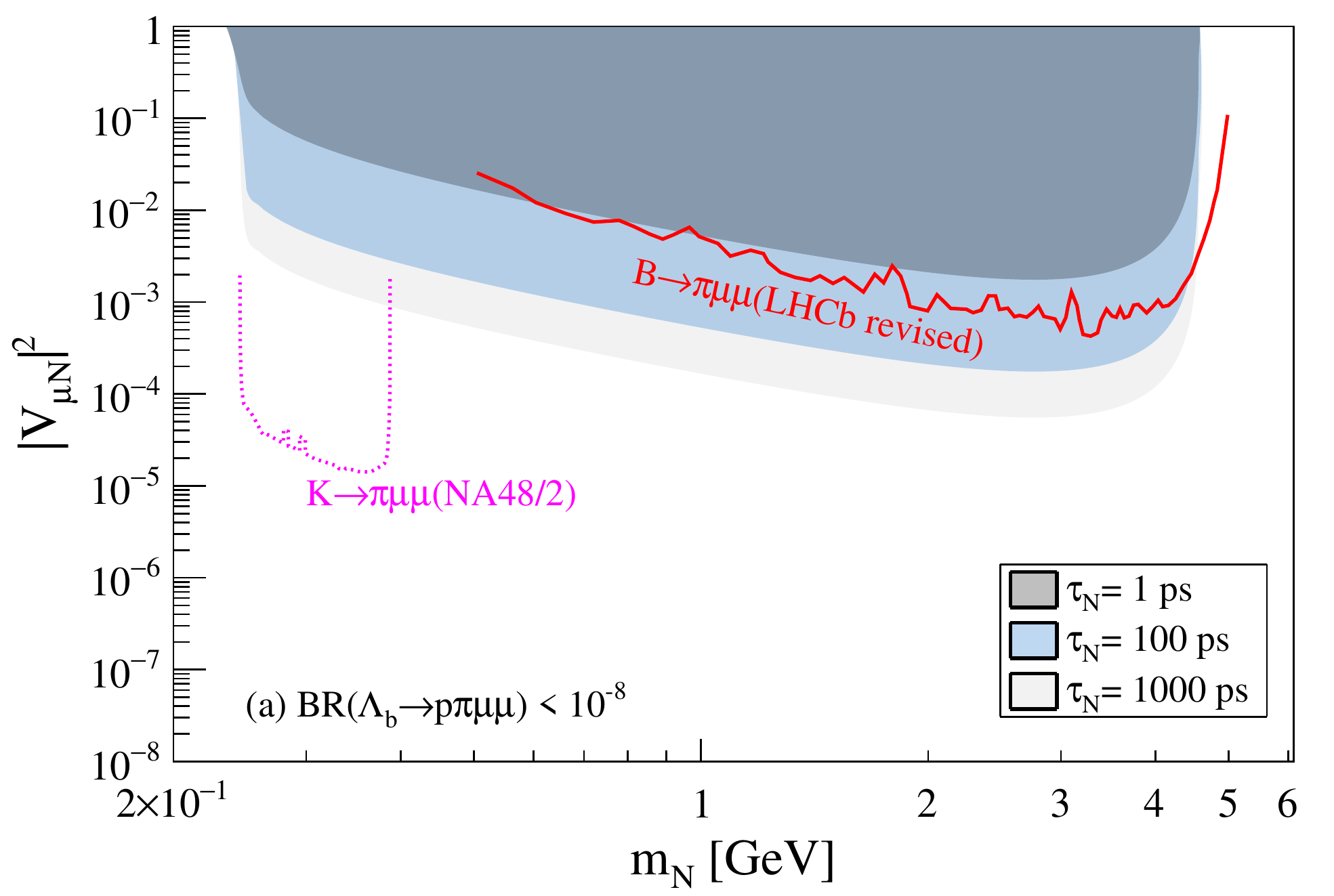}
\includegraphics[scale=0.44]{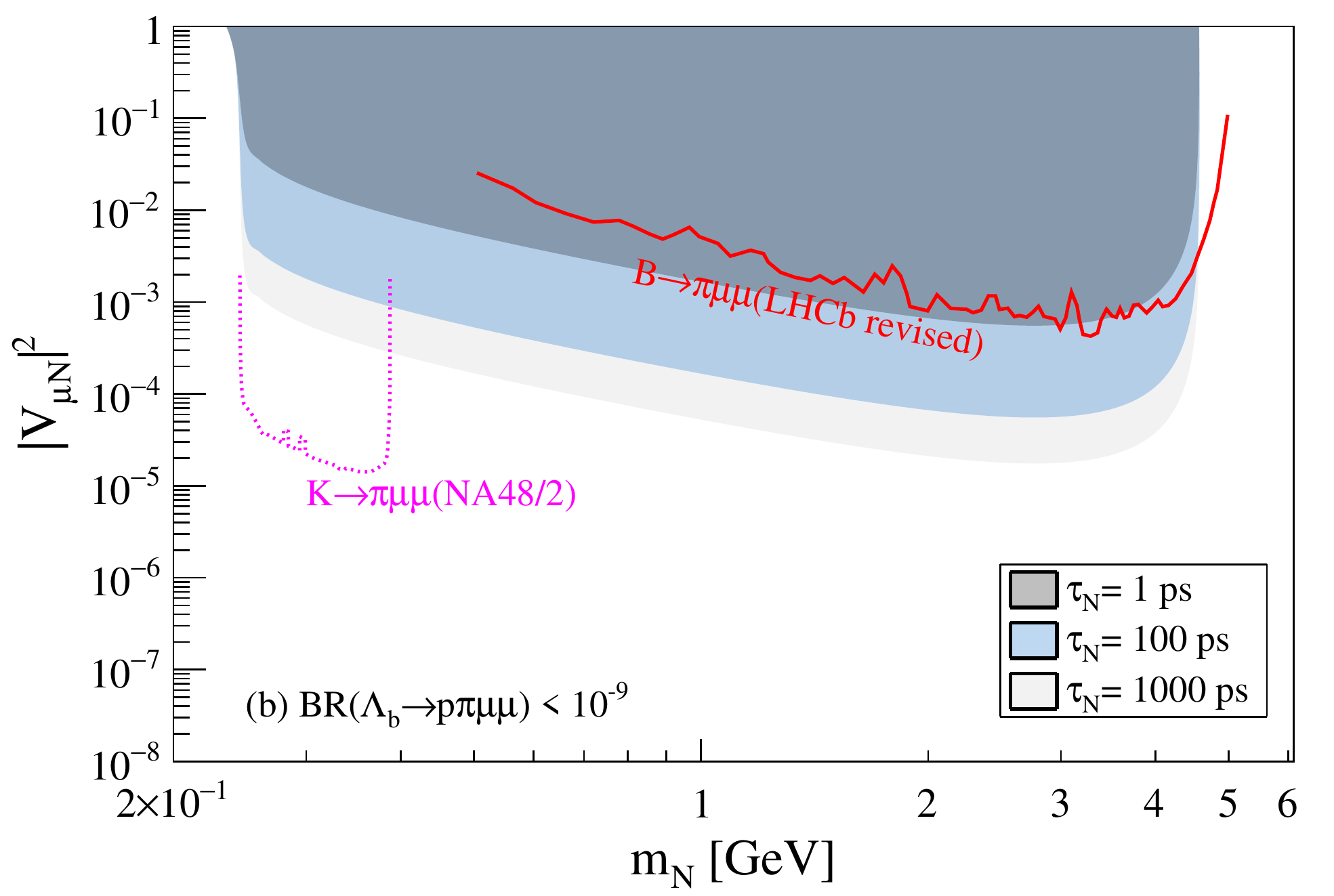}
\caption{\small Exclusion regions on the$(m_N, |V_{\mu N}|^{2})$ plane for (a) ${\rm BR}(\Lambda_b^0 \to p \pi^+\mu^- \mu^-) < 10^{-8}$ and  (b) ${\rm BR}(\Lambda_b^0 \to p \pi^+\mu^- \mu^-) < 10^{-9}$. The gray, blue, and light-gray regions represent the constraints obtained for heavy neutrino lifetimes of $\tau_N = [1, 100, 1000]$ ps, respectively. Limits provided by $K^- \rightarrow \pi^{+}\mu^{-}\mu^{-}$ \cite{CERNNA48/2:2016} and $B^{-} \to \pi^{+}\mu^{-}\mu^{-}$ \cite{Shuve:2016} are also included for comparison.}
\label{Fig4.1}
\end{figure}

To illustrate the constraints that can be achieved from the experimental searches on $\Lambda_b^0 \to  p \pi^+\mu^- \mu^-$, in Figs. \ref{Fig4.1}(a) and \ref{Fig4.1}(b), we show the exclusions regions on $|V_{\mu N}|^{2}$ as a function of $m_N$ obtained by taking an expected sensitivity on the branching fractions of the order ${\rm BR}(\Lambda_b^0 \to  p\pi^+\mu^- \mu^-) < 10^{-8}$ and $< 10^{-9}$, respectively.  In both cases, the gray, blue, light-gray regions represents the constraints obtained for heavy neutrino lifetimes of $\tau_N = [1, 100, 1000]$ ps, respectively. For the purpose of comparison, we also plotted the available exclusion limits obtained from searches on $|\Delta L|=2$ channels: $K^- \rightarrow \pi^{+}\mu^{-}\mu^{-}$ (NA48/2) \cite{CERNNA48/2:2016} and $B^{-} \to \pi^{+}\mu^{-}\mu^{-}$ (LHCb) \cite{LHCb:2014}. The limit from the $K^- \rightarrow \pi^{+}\mu^{-}\mu^{-}$ channel is taken for $\tau_N =$ 1000 ps \cite{CERNNA48/2:2016}. While for the $B^{-} \to \pi^{+}\mu^{-}\mu^{-}$ channel, we compare with the revised limit \cite{Shuve:2016} from the LHCb analysis \cite{LHCb:2014}.
We can see in Figs. \ref{Fig4.1}(a) and \ref{Fig4.1}(b) that the most restrictive constraint is given by $K^- \to \pi^+ \mu^-\mu^-$ which can reach $|V_{\mu N}|^2\sim \mathcal{O}(10^{-5})$ but only for a very narrow range $[0.25, 0.38]$ GeV of Majorana neutrino masses. For $m_N > 0.38$ GeV, the four-body channel $\Lambda_b^0 \to  p \pi^+\mu^- \mu^-$ (CKM suppressed) would be able to extend the region of $|V_{\mu N}|^{2}$ covered by the channel $B^- \to \pi^+\mu^-\mu^-$ (also CKM suppressed).

\begin{figure}[!t]
\centering
\includegraphics[scale=0.44]{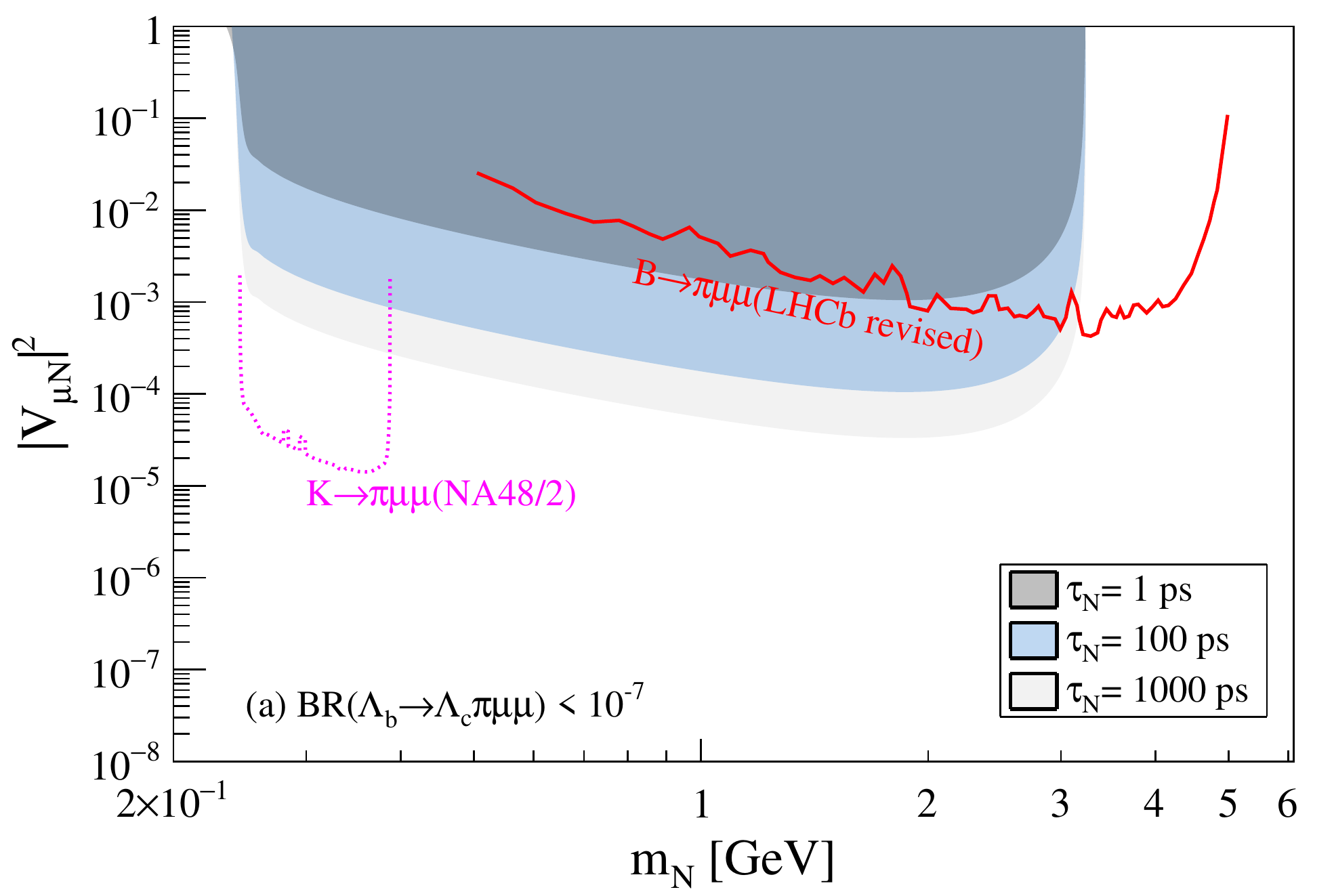}
\includegraphics[scale=0.44]{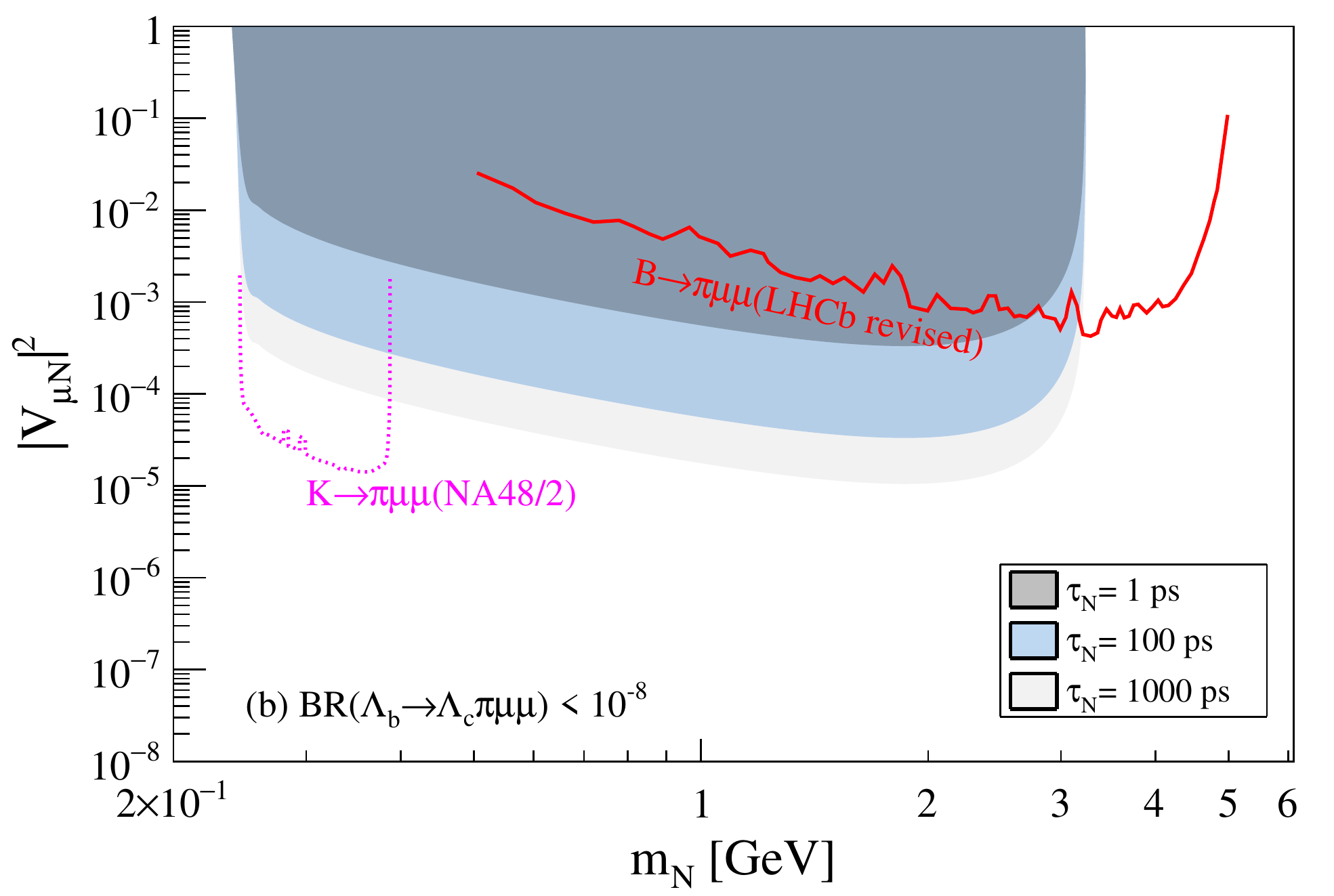}
\caption{\small Exclusion regions on the $(m_N, |V_{\mu N}|^{2})$ plane for (a) ${\rm BR}(\Lambda_b^0 \to \Lambda_c^+  \pi^+\mu^- \mu^-) < 10^{-7}$ and  (b) ${\rm BR}(\Lambda_b^0 \to \Lambda_c^+ \pi^+\mu^- \mu^-) < 10^{-8}$. The gray, blue, light-gray regiosn represent the constraints obtained for heavy neutrino lifetimes of $\tau_N = [1, 100, 1000]$ ps, respectively. Limits provided by $K^- \rightarrow \pi^{+}\mu^{-}\mu^{-}$ \cite{CERNNA48/2:2016} and $B^{-} \to \pi^{+}\mu^{-}\mu^{-}$ \cite{Shuve:2016} are also included for comparison.}
\label{Fig4.2}
\end{figure}

As for the searches on $\Lambda_b^0 \to  \Lambda_c^+ \pi^+\mu^- \mu^-$, with a similar expected sensitivity at the LHC of ${\rm BR}(\Lambda_b^0 \to  \Lambda_c^+ \pi^+\mu^- \mu^-) < 10^{-8}$ and $< 10^{-9}$, we show, respectively, the exclusion curves on the $(m_N,|V_{\mu N}|^2)$ plane in Figs. \ref{Fig4.2}(a) and \ref{Fig4.2}(b). Again, in both cases, the gray, blue, and light-gray regions present the constraints obtained for heavy neutrino lifetimes of $\tau_N = [1, 100, 1000]$ ps, respectively. It is important to remark that, due to the CKM mixing elements involved, the $\Lambda_b^0 \to  \Lambda_c^+ \pi^+\mu^- \mu^-$ channel is a CKM-allowed process, and therefore it would be able to exclude regions of $|V_{\mu N}|^{2}$ that are weaker than $K^- \to \pi^+\mu^-\mu^-$ and stronger than $B^- \to \pi^+\mu^-\mu^-$. 

In addition, it is important to comment that for the GeV-scale of sterile neutrino masses relevant to this work, $m_N \in [0.25,5.0]$ GeV, different search strategies have been used to get constraints on the mixing element $|V_{\mu N}|$ (for a recent review on the theoretical and experimental status see Refs.~\cite{Atre:2009,Deppisch:2015,Drewes:2013,Drewes:2015,deGouvea:2015,Fernandez-Martinez:2016} and references therein). The lack of experimental evidence of searches of peaks in the muon spectrum of leptonic $K^{\pm}$ decays (PS191, E949) and searches through specific visible channels of heavy neutrino decays produced in beam dump experiments (such as NA3, CHARM, and NuTeV, among others) allows us to put constraints on $|V_{\mu N}|^2 \sim \mathcal{O}(10^{-8} - 10^{-6})$  for masses of the  sterile neutrino ranging from 0.2 to 2.0 GeV \cite{Atre:2009,Deppisch:2015,Drewes:2013,Drewes:2015,deGouvea:2015}. It is expected that the recently proposed high-intensity beam dump experiment SHiP~\cite{SHiP} can significantly improve those bounds~\cite{Deppisch:2015}. Moreover, in the mass range [0.5,5.0] GeV searches of heavy neutrinos have been performed by Belle using the inclusive decay mode $B \to X \ell N$ followed by $N \to \ell\pi$ (with $\ell = e, \mu$) \cite{Belle:N}, and by DELPHI using the possible production of heavy neutrinos in the $Z$-boson decay $Z \to \nu N$ \cite{LEP}. For masses 5.0 GeV $<m_N < m_W$, the possibility of a heavy neutrino produced in $W$ and Higgs boson decays has been studied as well; see, for instance Ref.~\cite{mN_below_mW}.

\begin{figure}[!t]
\centering
\includegraphics[scale=0.45]{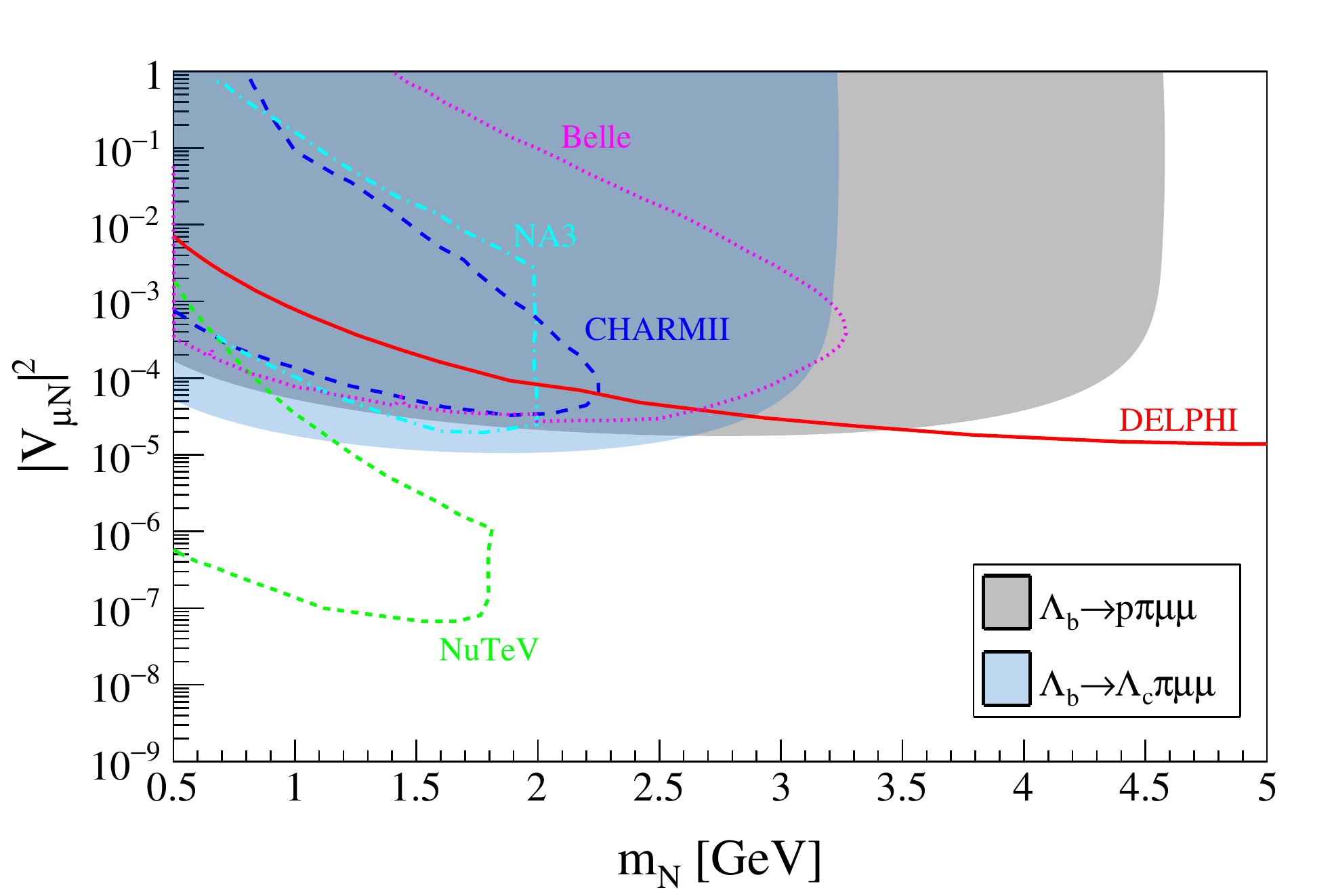}
\caption{\small Exclusion regions on $(m_N, |V_{\mu N}|^{2})$ plane coming from: Belle \cite{Belle:N}, DELPHI \cite{LEP}, NA3 \cite{NA3}, CHARMII \cite{CHARMII}, and NuTeV \cite{NuTeV} experiments. Limits provided by the searches on $\Lambda_b^0 \to (p,\Lambda_c^+) \pi^+\mu^- \mu^-$  are represented by the gray and blue regions, respectively. See text for details.}
\label{Fig4.3}
\end{figure}

In Fig. \ref{Fig4.3}, we show the exclusion bounds on the $(m_N,|V_{\mu N}|^2)$ plane coming from Belle~\cite{Belle:N}, DELPHI~\cite{LEP}, NA3~\cite{NA3}, CHARMII~\cite{CHARMII}, and NuTeV~\cite{NuTeV} experiments, in the mass range [0.5,5.0] GeV. In comparison, the constraints obtained from the searches on $\Lambda_b^0 \to (p,\Lambda_c^+) \pi^+\mu^- \mu^-$  are represented by the gray and blue regions, for a branching fraction of ${\rm BR} < 10^{-8}$ and ${\rm BR} < 10^{-9}$, respectively. In both cases, a lifetime $\tau_N = 1000$ ps have been taken as a representative value. We observe that our $\Delta L=2$ channels proposal would complement these bounds in the mass region around $m_N \simeq 2.0 - 3.0$ GeV.

\section{Conclusions} \label{Conclusion}

In the present work, we have studied new LNV processes in the four-body $|\Delta L|=2$ decays of the $\Lambda_b^0$ baryon, $\Lambda_b^0 \to p \pi^+\mu^- \mu^-$ and $\Lambda_b^0 \to \Lambda_c^+ \pi^+\mu^- \mu^-$. We have investigated the production of a GeV-scale Majorana neutrino through these $|\Delta L|=2$ decays, particularly, neutrino masses $m_N \in [0.25,4.57]$ GeV and $m_N \in [0.25,3.23]$ GeV. Because of the relatively high muon reconstruction system, we have paid attention on these same-sign dimuon channels and explored their experimental sensitivity at the LHCb and CMS. We considered heavy neutrino lifetimes of $\tau_N = [1, 100, 1000]$ ps, where the probability for the on-shell neutrino $N$ decay products to be inside the detector (acceptance factor $P_N$) has been taken into account in our analysis. According to this analysis, it is found as conservative values, that for a integrated luminosity collected of 10 and 50 fb${}^{-1}$ at the LHCb and 30, 300, and 3000 fb${}^{-1}$ at the CMS; one would expect sensitivities on the branching fractions of the order ${\rm BR}(\Lambda_b^0 \to p \pi^+\mu^- \mu^-) \lesssim \mathcal{O}(10^{-9} - 10^{-8})$ and  ${\rm BR}(\Lambda_b^0 \to \Lambda_c^+ \pi^+\mu^- \mu^-) \lesssim \mathcal{O}(10^{-8} - 10^{-7})$, respectively. With such sensitivities, we extracted constraints on the parameter space $(m_N,|V_{\mu N}|^2) $ that might be obtained from their experimental search for neutrino lifetimes of $\tau_N = [1, 100, 1000]$ ps. Depending on the $\tau_N$ value, these channels would be able to exclude regions of $|V_{\mu N}|^{2}$  that are weaker than $K^- \to \pi^+\mu^-\mu^-$ (NA48/2)  and stronger than $B^- \to \pi^+\mu^-\mu^-$ (LHCb). In addition, we observed that our $\Delta L=2$ channels proposal would complement the bounds given by different search strategies (such as NA3, CHARMII, NuTeV, Belle, and DELPHI), in the mass region around $m_N \simeq 2.0 - 3.0$ GeV. Consequently, the study of  $\Delta L=2$ decays of the $\Lambda_b$ baryon is a very promising place to look for heavy Majorana neutrinos.

\acknowledgments

The author N. Quintero acknowledges support from Direcci\'{o}n General de Investigaciones - Universidad Santiago de Cali under Project No. 935-621717-016. J. D. Ruiz-\'{A}lvarez gratefully acknowledges the support of COLCIENCIAS, the Administrative Department of Science, Technology and Innovation of Colombia. The work of J. Mej\'{i}a-Guisao has been financially supported by Conacyt (M\'exico) under Projects No. 296 (Fronteras de la
Ciencia), No. 221329, and No. 250607 (Ciencia B\'{a}sica).










\end{document}